\documentclass[conference]{IEEEtran}
\IEEEoverridecommandlockouts

\usepackage{amssymb}
\usepackage[cmex10]{amsmath}

\usepackage{stfloats}
\usepackage{graphicx}
\usepackage{subfigure}
\usepackage{tabularx}
\usepackage{epsfig,epsf,color,balance,cite}
\usepackage{verbatim}
\usepackage{url}
\usepackage{bm}
\usepackage[utf8]{inputenc}
\usepackage{tikz}
\usepackage{tkz-graph}
\usetikzlibrary{positioning,chains,fit,shapes,calc}

\usepackage{algorithm}
\usepackage{algorithmic}

\usepackage{color}
\definecolor{myc1}{rgb}{0,0,0}

\begin{document}

% paper title
\title{{Optimized Asymmetric Feedback Detection for Rate-adaptive HARQ with Unreliable Feedback}  }

\author{
\IEEEauthorblockN{Weihang Ding\IEEEauthorrefmark{1} and Mohammad Shikh-Bahaei\IEEEauthorrefmark{1}
                  }
\IEEEauthorblockA{\IEEEauthorrefmark{1}Centre for Telecommunications Research, Department of Engineering, King's College London, London WC2R 2LS, UK.}

\vspace{-2em}
}

%\author{
%\IEEEauthorblockN{Zhaohui Yang\IEEEauthorrefmark{1},
%                  Mingzhe Chen\IEEEauthorrefmark{2},
%                  Walid Saad\IEEEauthorrefmark{3},
%                  Wei Xu\IEEEauthorrefmark{4},
%                  and Mohammad Shikh-Bahaei\IEEEauthorrefmark{1}
%                  }
%\IEEEauthorblockA{\IEEEauthorrefmark{1}Centre for Telecommunications Research, Department of Informatics, King’s College London, WC2B 4BG, UK, Emails: yang.zhaohui@kcl.ac.uk, m.sbahaei@kcl.ac.uk.}
%\IEEEauthorblockA{\IEEEauthorrefmark{2}Beijing Key Laboratory of Network System Architecture and Convergence,
%Beijing University of Posts and Telecommunications, Beijing, China 100876, Email: chenmingzhe@bupt.edu.cn.}
%\IEEEauthorblockA{\IEEEauthorrefmark{3}Wireless@VT, Bradley Department of Electrical and Computer Engineering, Virginia Tech, Blacksburg, VA, USA, Email: walids@vt.edu.}
%\IEEEauthorblockA{\IEEEauthorrefmark{2}National Mobile Communications Research Laboratory, Southeast University, Nanjing 211111, China, Email: wxu@seu.edu.cn.}
%}
% make the title area
\maketitle

\begin{abstract}
This work considers downlink incremental redundancy Hybrid Automatic Repeat Request (IR-HARQ) over unreliable feedback channels. Since the impact of positive feedback (i.e., ACK) error is smaller than that of negative feedback (i.e., NACK) error, an asymmetric feedback detection scheme is proposed to protect NACK and further reduce the outage probability. We formulate the HARQ process as a Markov Decision Process (MDP) model to adapt to the transmission rate of each transmission attempt without enriched feedback and additional feedback cost.
We aim to optimize the performance of HARQ process under certain outage probability requirements by finding optimal asymmetric detection thresholds. 
Numerical results obtained on the downlink Rayleigh fading channel and 5G new radio (NR) PUCCH feedback channel show that by applying asymmetric feedback detection and adaptive rate allocation, higher throughput can be achieved under outage probability limitations.

\end{abstract}

\begin{IEEEkeywords}
Hybrid automatic repeat request, rate adaptation, asymmetric feedback detection
\end{IEEEkeywords}
\IEEEpeerreviewmaketitle

\section{Introduction}

Hybrid Automatic Repeat Request (HARQ) is a mechanism combining Forward Error Correction (FEC) and Automatic Repeat Request (ARQ). It operates on the physical layer and MAC sublayer, handling retransmissions of missing or erroneous data.
Generally, there are two types of soft combining methods in HARQ, namely chase combining (CC-HARQ) and incremental redundancy (IR-HARQ) \cite{Caire}. In CC-HARQ, the same packet is retransmitted and the receiver combines the bits received in different transmissions using maximum-ratio combining, which results in a $E_b/N_0$ gain.
In IR-HARQ, additional redundancy is retransmittted and extra coding gain is obtained. In this paper, we only focus on IR-HARQ because it is more powerful and more commonly used in modern communication systems.

In the conventional IR-HARQ protocol, the length and power of each transmission is fixed in all HARQ rounds. Binary feedback is used for informing the decoding status at the receiver\cite{Jabi}. If ACK is observed by the transmitter, the transmission is believed to be successful. In contrast, NACK indicates a failed transmission.
A disadvantage is that some resources will be wasted when NACK is observed because the transmitter does not know how much additional redundancy is actually required \cite{Trillingsgaardems}.

Application of adaptive HARQ can reduce this waste of resources because the transmitter can adaptively choose the optimal strategy for the next transmission. Adaptive HARQ can include rate adaptation \cite{Szczecinski, Khosrvirad2014, My2}, power adaptation \cite{Tuninetti}, \cite{Chaitanya}, and adaptive modulation and coding (AMC) schemes\cite{Choi, Yirun1, Nehra, Bobarshad, PCF, My1}.

The majority of adaptive HARQ problems are solved based on the assumption that the decoding state information (DSI) is perfectly transmitted to the transmitter through feedback, which is practically impossible because of deviations when discretization is performed. Moreover, the feedback channel is usually more costly, hence it is impractical to use very long feedback. When the quality of the feedback channel is low, applying extra feedback bits will increase the feedback error rate and make the error pattern harder to analyse.
In \cite{Jabi}, the relationship between feedback resolution and overall throughput is studied. The results show that if the feedback includes more than 8 bits, the performance is very close to an ideal feedback HARQ system.

Because of high cost, the error probability of the HARQ feedback can not be made arbitrarily low. A feedback error rate of 1\% is reasonable in LTE \cite{LTEchap12} and 0.1\%-1\% in 5G NR\cite{5GNR}. Unreliable feedback will impair the performance of the HARQ system. There are only a few contributions on HARQ with unreliable feedback. In \cite{Breddermann2014}, \cite{Ahmad}, the feedback channel is modeled as a Binary Symmetric Channel (BSC). In \cite{Malak}, a notion of guaranteeable delay is introduced while the feedback channel is an erasure channel. Different from conventional symmetric feedback detection, asymmetric feedback detection is introduced in \cite{Shariatmadari} to provide a better protection to NACK without assigning extra resources. However, asymmetric detection with variable thresholds has never been analysed in the area of HARQ before. 

The major contributions of this work are:
\begin{itemize}
    \item We introduce a rate-adaptive scheme for single-bit feedback HARQ process. The rate adaptation problem is modeled as a Markov Decision Process (MDP). We show how to use dynamic programming to optimize the throughput of this HARQ process under strict outage probability limitations.
    \item We analyse the throughput degradation of HARQ system caused by unreliable feedback. We propose an asymmetric feedback detection scheme to enhance the HARQ performance under unreliable feedback channel.
    \item In the case of variable thresholds, we formulate a performance optimization problem to maximize HARQ throughput under certain outage probability limitation.
\end{itemize}

The rest of this paper is organized as follows. In Section II, we introduce the system model of both data and feedback channel. In Section III and IV, we analyze the HARQ performance under reliable and unreliable feedback channel and propose the optimization problem. The nonconvex optimization problem is solved in Section V via an iterative algorithm. Simulation
results are shown in Section VI and conclusions are finally
drawn in Section VII.

\section{System model}
All the packets from higher layers are segmented into blocks of length $N_b$. After encoding, the code can be transmitted with arbitrary rate. In the first transmission attempt, $N_1$ symbols are transmitted initially. $N_2,\dots,N_M$ symbols are transmitted in the following transmission attempts where $M$ is the maximum allowed number of transmissions of this HARQ process.
If the decoding process at the receiver fails, a NACK will be released, which will notify the transmitter to transmit a certain amount of redundancy in the subsequent transmission attempt. This continues until the number of transmissions reaches $M$, or the transmitter observes an ACK, which indicates a successful decoding.
The uplink channel is more likely to be unreliable due to the energy limitation on the battery of user equipment. In downlink HARQ, the feedback channel is uplink. Hence, we merely focus on downlink HARQ in this paper.

\subsection{Downlink channel}
The data is transmitted over a single-user block-fading channel under Gaussian noise. The channel remains constant over a single transmission but are independent from one transmission to another. The $k$-th received signal is given by:
\begin{equation}
    \bm{y}_k=\sqrt{\text{SNR}_d}h_k\bm{x}_k+\bm{z}_k
\end{equation}
where $\bm{x_k}$, $\bm{y_k}$, and $\bm{z_k}$ are the $k$-th transmitted symbol, received symbol and additive noise respectively, $h_k$ is the channel fading coefficient, SNR$_d$ is the average received signal to noise ratio of the downlink channel. 

We assume that the fading channel varies rapidly, hence $h_k$ can be seen as independent identically distributed (i.i.d) variables and remains constant during a single transmission attempt. Hence, the transmitter does not know $h_k$ and cannot predict it using the previous feedback prior to transmission. $|h_k|$ is Rayleigh distributed with normalized unity power gain $\mathbb{E}[|h_k|^2]=1$. $\bm{z}_k$ is a complex vector of zero mean, and unitary-variance Gaussian variables representing the noise. It is assumed that the transmitted symbol $\bm{x_k}$ has unit power and SNR$_d$, which is known by the transmitter prior to the data transmission, remains constant in each HARQ round.

In actual communication systems, the above assumptions are reasonable. Although we only consider Rayleigh fading channel and single antenna system, it can be easily extended to other channels and MIMO systems.

\subsection{Uplink channel}
In 5G NR, Physical uplink control channel (PUCCH) is used for transmitting uplink control messages. There are 5 PUCCH formats in 5G NR \cite{38.211}, among which PUCCH format 0 and PUCCH format 1 are commonly used for HARQ ACK/NACK. In this paper, we use PUCCH format 0 with 1 orthogonal frequency-division multiplexing (OFDM) symbol only.

PUCCH format 0 is a short PUCCH format which can transmit 1-2 bits over one or two OFDM symbols based on sequence selection. Two different sequences are selected for information bit '1' (ACK) and '0' (NACK) if 1-bit information is transmitted. These sequences are generated by phase rotation of the same base sequence\cite{5GNR}. Each OFDM symbol contains 12 orthogonal symbols corresponding to the 12 sub-carriers in the frequency domain contained in the resource block. 
If the PUCCH contains only one bit of HARQ feedback, 6 of the total 12 elements in the sequence are different. The phase difference of each of these 6 symbol pairs is $\pi$.

\section{HARQ performance}
 The decoding is based on the received sequence $\bm{y}_k$, and uses maximum likelihood decoding scheme.
The overall accumulated mutual information (MI) at the receiver in IR-HARQ is the summation of MI over all transmission attempts \cite{Jabi}. 

To simplify, like \cite{Szczecinski} we define a normalized index $\rho_k=\frac{N_k}{N_b},k=1,2,\dots,M$, which indicates how many symbols are used to transmit each information bit in the $k$-th transmission. The probability that the decoding fails after the $k$-th transmission attempt $P_{k,f}$ can be written as\cite{Tse}:
\begin{equation}
\begin{aligned}
    P_{k,f}&=\mathbb{P}\left\{\sum_{l=1}^{k}I_l< 1\right\}\\
\end{aligned}
\end{equation}
where $I_l$ is the normalized MI in the $l$-th transmission defined as:
\begin{equation}
    I_l = \rho_l \log (1+|h_l|^2\text{SNR}_d)
\end{equation}

A general criterion to evaluate the performance of HARQ processes is the throughput, which is the ratio of the number of information bits correctly transmitted over the number of channel uses. The throughput $\eta$ of this HARQ process can be written as:
\begin{equation}
    \eta=\frac{N_{b}}{\mathbb{E}[N_{s}]}(1-P_{out})
\end{equation}
where $\mathbb{E}[N_{s}]$ is the average number of channels required to transmit one block and $P_{out}$ is the outage probability of this HARQ process.

If the HARQ feedback channel can be regarded as reliable, an outage will only occur when the decoder is still not able to decode a block after the number of transmissions reaches the maximum limit, i.e., the decoding still fails after $M$ transmissions. The outage probability after the $M$-th transmission can be written as:
\begin{equation}
    P_{out}=P_{M,f}=\mathbb{P}\left\{\sum_{k=1}^{M}I_k< 1\right\}
\end{equation}

To calculate $P_{k,f}$, we need to find the probability density function (pdf) of the accumulated MI. Let $Z_k=\sum_{l=1}^{k} I_l$ denote the summation of $I$ in the first $k$ transmissions.
The outage probability after the $k$-th transmission can be calculated as
\begin{equation}
    P_{k,f}=\int_{0}^{1}f(Z_k)dZ_k
    \label{e1}
\end{equation}
where $f(Z_k)$ is the pdf of $Z_k$ computed through the convolution of $f(I_1), \dots, f(I_k)$:
\begin{equation}
    f(Z_k) = f(I_1)*\dots *f(I_k)
    \label{e2}
\end{equation}

The throughput of such a HARQ process can be further written as:
\begin{equation}
    \eta=\frac{1-P_{out}}{\sum_{i=1}^{M}\rho_iP_{i-1,f}}
\end{equation}
where $P_{0,f}=1$ for initialization.

\section{HARQ with unreliable feedback}

In 5G NR, HARQ handles retransmission of missing or erroneous data with the supplement of RLC layer protocols. Because of the two-level retransmission structure, it is unnecessary to eliminate the HARQ failure rate to 0. The transmission power is always limited by the battery capacity of the mobile devices. In order to keep the feedback cost reasonable, a feedback error rate of around 0.1\%-1\% is acceptable.

\begin{figure}[!t]
	\centering\includegraphics[width=0.7\columnwidth]{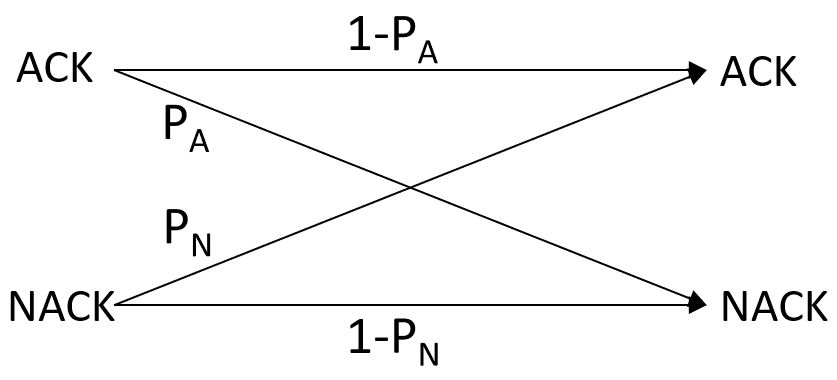}
	\caption{A general model of unreliable HARQ feedback channel}
	\label{asymmetric}
\end{figure}

If an ACK$\to$NACK error occurs, the transmitter will transmit certain amount of redundancy in the next time slot which will result in a small waste of throughput. However, if the transmitter recognizes an NACK as ACK, it will cause an outage and significant degradation on the throughput and delay. Obviously, the impact of NACK$\to$ACK error is much greater than ACK$\to$NACK error, so the protection on NACK is more important than ACK. When the feedback channel resources are strictly limited, in order to provide better protection on NACK, the protection on ACK can be sacrificed. 
A general model of unreliable HARQ feedback channel is depicted in Fig.\ref{asymmetric}. This channel model is defined by two conditional error probabilities $P_A$ and $P_N$ indicating the error rate of ACK and NACK respectively.

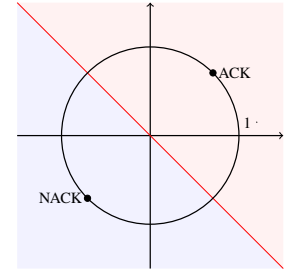
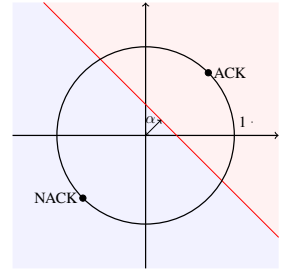
\begin{figure}[!t]
\centering
\subfigure[Symmetric detection]{
    \resizebox{0.2\textwidth}{!}{%
    \begin{tikzpicture}
    \fill [blue!5](-3,3) -- (-3,3) -- (-3,-3) -- (3,-3) -- (3,-3);
    \fill [red!5](3,3) -- (-3,3) -- (3,-3);
    \draw [thick,->](-3,0) -- (3,0);
    \draw [thick,->](0,-3) -- (0,3);
    \draw [thin,red](-3,3) -- (3,-3);
    \draw [thick] (0,0) circle (2);
    \filldraw[black] (1.41421,1.41421) circle (2pt) node[anchor=west] {ACK};
    \filldraw[black] (-1.41421,-1.41421) circle (2pt) node[anchor=east] {NACK};
    \filldraw[black] (2.4,0.3) circle (0.1pt) node[anchor=east] {1};
    \end{tikzpicture}
    }
    \label{PSKsym}
}

\subfigure[Asymmetric detection]{
    \resizebox{0.2\textwidth}{!}{%
    \begin{tikzpicture}
    \fill [blue!5](-2.3,3) -- (-3,3) -- (-3,-3) -- (3,-3) -- (3,-2.3);
    \fill [red!5](3,3) -- (-2.3,3) -- (3,-2.3);
    \draw [thick,->](-3,0) -- (3,0);
    \draw [thick,->](0,-3) -- (0,3);
    \draw [thin,red](-2.3,3) -- (3,-2.3);
    \draw [thick] (0,0) circle (2);
    \draw [thin,<->] (0,0) -- (0.35,0.35) node[anchor=east] {$\alpha$};
    \filldraw[black] (1.41421,1.41421) circle (2pt) node[anchor=west] {ACK};
    \filldraw[black] (-1.41421,-1.41421) circle (2pt) node[anchor=east] {NACK};
    \filldraw[black] (2.4,0.3) circle (0.1pt) node[anchor=east] {1};
    \end{tikzpicture}
    }
    \label{Asym sch1}
}
\caption{Two HARQ feedback detection schemes.}
\label{Asym}
\end{figure}

If the detection threshold is set to be the same as the red line in Fig.\ref{PSKsym}, it is symmetric detection. In this case, the area of ACK and NACK on the constellation are the same so that ACK and NACK error rates are equal.
Now, we consider asymmetric feedback detection to decrease NACK$\to$ACK error rate. 
To eliminate NACK$\to$ACK error rate, the area of NACK should be larger than ACK on the constellation. We propose an asymmetric feedback detection schemes to match the OFDM symbol to ACK and NACK, which is illustrated in Fig.\ref{Asym sch1}. $\alpha \in \mathbb{R}$ is an asymmetric detection index defined by the ratio of the minimum Euclidean distance from the asymmetric decoding threshold to the origin of the constellation diagram over the amplitude. The amplitude of the signal is normalized in Fig.\ref{Asym}.

\newcounter{mytempeqncnt}
\begin{figure*}[!t]
\normalsize
\setcounter{mytempeqncnt}{\value{equation}}
\begin{equation}
    P_{out}^{[un]}=1-\left(1-P_{N,1}P_{1,f}-\sum_{i=2}^{M-1}\left(P_{N,i}P_{i,f}\prod_{j=1}^{i-1}(1-P_{N,j})\right)\right)(1-P_{out})
    \label{Pout}
\end{equation}
\vspace{-.5em}
\begin{equation}
P_i=\left\{
\begin{aligned}
    &1,\:\:\:\:\:\:\:\:\text{if i=1}\\
    &P_{i-1,f}\prod_{j=1}^{i-1}(1-P_{N,j})+\sum_{j=1}^{i-1}\left((P_{i-j-1,f}-P_{i-j,f})\prod_{k=0}^{i-j-1}(1-P_{N,k})\prod_{k=i-j}^{i-1}P_{A,k}\right),\:\:\:\:\:\text{if i=2\dots M}\\
\end{aligned}
\right.
\label{Pi}
\end{equation}
\hrulefill
\vspace*{-4pt}
\vspace{-1em}
\end{figure*}

The asymmetric detection index $\alpha_i$ after the $i$-th transmission attempt depends largely on the probabilities of ACK and NACK. Because 6 of the total 12 elements in the sequence are opposite, if the feedback channel is an additive white Gaussian noise (AWGN) channel, the NACK/ACK error rate after of the $i$-th feedback can be respectively written as:
\begin{equation}
    P_{N,i}=\frac{1}{2}erfc\left((1+\alpha_i)\sqrt{6\text{SNR}_u}\right)
\end{equation}
\begin{equation}
    P_{A,i}=\frac{1}{2}erfc\left((1-\alpha_i)\sqrt{6\text{SNR}_u}\right)
\end{equation}
where SNR$_u$ is the signal-to-noise ratio of uplink feedback channel and $\alpha_i$ is the asymmetric detection index of the $i$-th feedback.

The outage probability of this unreliable feedback HARQ process can be written as (\ref{Pout}), which is at the top of the next page. The outage probability of the HARQ process is limited by both the uplink and downlink channel and is lower bounded by the outage probability of the HARQ process with perfect feedback channel.

The average number of symbols transmitted in the whole HARQ round can be calculated as:
\begin{equation}
\mathbb{E}[N_{s}^{[un]}]=\sum_{i=1}^{M}\rho_iN_bP_i
\end{equation}
where $P_i$ is the probability that the $i$-th transmission occurs and can be calculated through (\ref{Pi}).

Our goal is to maximize the overall throughput of the HARQ process in the presence of outage probability constraints by selecting optimal $\rho_1,\dots,\rho_M$, and $\alpha_1,\dots,\alpha_{M-1}$. This optimization problem can be formulated as:
\begin{equation}
\begin{aligned}
    \max_{\rho_1,\dots,\rho_M,\alpha_1\dots \alpha_{M-1}}&\frac{N_{b}}{\mathbb{E}[N_{s}^{[un]}]}(1-P_{out}^{[un]})\\
    \rm{subject\:to:}&\:\:P_{out}^{[un]}\leq \epsilon\\
    &\:\:\sum_{k=1}^{M}\rho_k \leq\frac{N_m}{N_b}\\
    &\:\:\rho_{min}\leq \rho_k \leq \rho_{max},\:\:k=1,2,\dots,M\\
    &\:\:\alpha_k \in \mathbb{R},\:\:k=1,2,\dots,M-1\\
\end{aligned}
\label{opt}
\end{equation}
where $N_m$ is the length of the mother codeword, $\rho_{min}$ and $\rho_{max}$ are the lower and upper bound of normalized transmission rate in each transmission attempt respectively, $\epsilon$ is the outage probability limit which is usually 0.1-1\% for normal HARQ process\cite{5GNR}. Problem (\ref{opt}) is a nonconvex problem so it is generally hard to effectively obtain a globally optimal solution. 

\section{Proposed algorithm}

In this section, in order to solve problem (\ref{opt}) we propose an algorithm to iteratively optimize $\rho_1,\dots,\rho_M$ with fixed $\alpha_1,\dots,\alpha_{M-1}$, and optimize $\alpha_1,\dots,\alpha_{M-1}$ with updated fixed $\rho_1,\dots,\rho_M$.

\subsection{Optimize transmission rates}
Problem (\ref{opt}) with given $\alpha_1,\dots,\alpha_{M-1}$ can be rewritten as:

\begin{equation}
\begin{aligned}
\max_{\rho_1,\dots,\rho_M}&\frac{N_{b}}{\mathbb{E}[N_{s}^{[un]}]}(1-P_{out}^{[un]})\\
\rm{subject\:to:}\:\:&P_{out}^{[un]}\leq \epsilon\\
&\sum_{k=1}^{M}\rho_k \leq\frac{N_m}{N_b}\\
&\rho_{min}\leq \rho_k \leq \rho_{max},\:\:k=1,2,\dots,M
\end{aligned}
\label{op1}
\end{equation}

This problem can be formulated as a MDP framework and can be solved using dynamic programming. In order to use MDP and dynamic programming techniques, the length of the transmitted codeword needs to be discretized at first.
It is hard to obtain the outage probability after the $k$-th transmission via (\ref{e1}) and (\ref{e2}) because convolution is involved. The accumulated MI can be approximated by a Gaussian variable which is accurate over low and moderate SNR regimes\cite{Wu}. The approximated pdf of accumulated MI $Z_k$ can be written as:

\begin{equation}
    f(Z_k) = \frac{1}{\sqrt{2\pi\sum_{i=1}^{k}\rho_i^2\sigma_I^2}}e^{-\frac{(Z_k-\sum_{i=1}^{k}\rho_i\bar{I})^2}{2\sum_{i=1}^{k}\rho_i^2\sigma_I^2}}
    \label{pdf}
\end{equation}
where $\bar{I}$ is the mean value of MI and $\sigma_I^2$ is the variance of MI given by \cite{McKay}:

\begin{equation}
    \bar{I}=\log_2(e)e^{\frac{1}{\text{SNR}_d}}\int_{1}^{\infty}t^{-1}e^{-\frac{t}{\text{SNR}_d}}dt
\end{equation}

\begin{equation}
    \begin{aligned}
    \sigma_I^2=\frac{2}{\text{SNR}_d}\log_2^2(e)e^{\frac{1}{\text{SNR}_d}} G^{4,0}_{3,4}\left(1/\text{SNR}_d|_{0,-1,-1,-1}^{0,0,0}\right)-\bar{I}^2
    \end{aligned}
\end{equation}
where $G^{m,n}_{p,q}\left(^{a_1,\dots,a_p}_{b_1,\dots,b_q}|z\right)$ is the Meijer G-function.

The outage probability after the $k$-th transmission attempt can be written as:

\begin{equation}
    P_{k,f}=\int_{0}^{1}f(z_k)dZ_k\approx Q\left(\frac{\sum_{i=1}^{k}\rho_i\bar{I}-1}{\sqrt{\sum_{i=1}^{k}\rho_i^2}\sigma_I}\right)
\end{equation}
where $Q(\cdot)$ is the tail distributed function of the standard normal distribution.

The probability that an outage occurs after the $k$-th transmission attempt $P_{out,k}$ can be written as:

\begin{equation}
P_{out,k}=\left\{
\begin{aligned}
&P_{N,1}P_{1,f},\:\:\:\:\:\:\:\:\text{if k=1}\\
&\frac{P_{N,1}P_{1,f}+\sum_{i=2}^{k}\left(P_{N,i}P_{i,f}\prod_{j=1}^{i-1}(1-P_{N,j})\right)}{P_k},\\
&\:\:\:\:\:\:\:\:\:\:\:\:\:\:\:\:\:\:\:\:\:\:\:\:\:\:\:\text{if k=2,\dots,M-1}\\
&P_{k,f}/P_k,\:\:\:\:\:\:\:\:\:\text{if k=M}\\
\end{aligned}
\right.
\end{equation}

Based on problem (\ref{op1}), we formulate a dual problem using Lagrangian $\mathcal{L}$:

\begin{equation}
\begin{aligned}
    \mathcal{L}(\lambda)=&\min_{\rho_1,\dots,\rho_M\in \mathbb{R}}\sum_{i=1}^{M}\rho_iP_i+\lambda P_{out}^{[un]}
\end{aligned}
\end{equation}

According to (\ref{Pi}) and (\ref{pdf}), the states of the Markov chain should include the discretized rate of all previous transmission attempts, so the other constraints in problem (\ref{op1}) are not needed to consider in the dual problem. Normally, $M$ is not a very large value, so it is practical to solve this MDP framework using dynamic programming. The dynamic programming recursions can be written as:

\begin{equation}
    V(S_0)=\min_{\rho_{min}\leq\rho_1\leq \rho_{max}} V(\rho_1) 
\end{equation}

\begin{equation}
    V(\rho_1)=\min_{\rho_{min}\leq\rho_2\leq \rho_{max}} \rho_1+\frac{P_2}{P_1}V(\rho_1,\rho_2)+\lambda P_{out,1}
\end{equation}

\begin{equation*}
    \dots
\end{equation*}

\begin{equation}
\begin{aligned}
    V(\rho_1,\dots,&\rho_{M-1})=\min_{\rho_{min}\leq\rho_M\leq \rho_{max}} \rho_{M-1}\\
    &+\frac{P_M}{P_{M-1}}V(\rho_1,\dots,\rho_M)+\lambda P_{out,M-1}
\end{aligned}
\end{equation}

\begin{equation}
    V(\rho_1,\dots,\rho_M)=\rho_M+\lambda P_{out,M}
\end{equation}

It is quite obvious that when the throughput is maximized, the obtained outage probability is a monotonically non-increasing function of $\lambda$.
Thus, the above problem can be solved by finding $\lambda$ such that $P_{out}^{[un]}=\epsilon$.

\subsection{Optimize asymmetric detection thresholds}

Similarly, problem (\ref{opt}) with given $\rho_1,\dots,\rho_M$ can be rewritten as:

\begin{equation}
\begin{aligned}
    \max_{\alpha_1\dots \alpha_{M-1}\in \mathbb{R}}&\frac{N_{b}}{\mathbb{E}[N_{s}^{[un]}]}(1-\epsilon)\\
    \rm{subject\:to:}&\:\:P_{out}^{[un]}\leq \epsilon\\
\end{aligned}
\label{op2}
\end{equation}

Since throughput $\eta$ is monotonically decreasing with increasing $\alpha$, the constrained optimization problem (\ref{op2}) can be directly solved using Projected Gradient Descent (PGD) technique.

\subsection{Iterative algorithm}

By iteratively solving problem (\ref{op1}) and problem (\ref{op2}), the optimal code rates and asymmetric thresholds can be obtained using Algorithm \ref{Alg1}. Since the optimal result is obtained in each step, the objective value of problem (\ref{opt}) is non-decreasing in each iteration. Moreover, the throughput of this HARQ process is upper bounded by the ergodic capacity of the downlink channel. Therefore, Algorithm \ref{Alg1} must converge.

\begin{algorithm}[t]
\algsetup{linenosize=\small}
\small
\caption{Iterative code rate and asymmetric threshold optimization}
\begin{algorithmic}[1]
 \STATE Set the initial solution $\rho_1,\dots,\rho_M$, $\alpha_1,\dots,\alpha_{M-1}$ of problem (\ref{opt}). Set the iteration number $n=0$.
 \REPEAT
 \STATE Obtain the optimal solution of problem (\ref{op1}) with fixed $\alpha_1,\dots,\alpha_{M-1}$.
  \STATE Obtain the optimal solution of problem (\ref{op2}) with fixed $\rho_1,\dots,\rho_M$.
  \STATE Set $n=n+1$.
 \UNTIL{the objective value converges}
\end{algorithmic}
\label{Alg1}
\end{algorithm}

\section{Numerical results}
For the simulation, the mother code is separated into 64 minimum transmission units. The maximum number of transmissions $M$ is 4. We set $M=4$ because allowing more transmissions will cause too much delay, and there are 4 redundancy versions in 5G LDPC codes.
Since a feedback error rate of 1\% is reasonable in LTE and 0.1\%-1\% is acceptable in 5G NR, the outage probability limit $\epsilon$ is assumed to be 1\%. Generally, the comparison of different HARQ processes is not straightforward. In this work, we use throughput as the only criterion for evaluating the performance of the HARQ process.

\begin{figure}[!t]
\centering
\includegraphics[width=\linewidth]{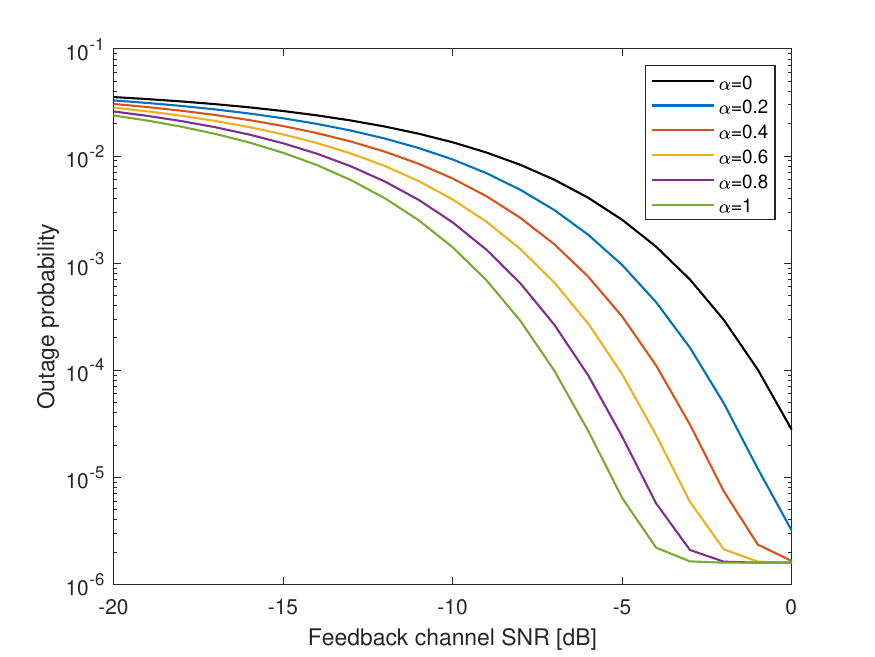}
\centering
\caption{Minimum achievable HARQ outage probability when SNR$_d$=3dB}
\label{alpha with rate adaptation}
\end{figure}

The outage probability of HARQ with asymmetric feedback detection is shown in Fig.\ref{alpha with rate adaptation}. Here, $\alpha$ is fixed among all transmission attempts. The outage probabilities of HARQ with asymmetric detection for $\alpha=0.2,0.4,0.6,0.8,1$ are compared with conventional symmetric feedback detection.
It is obvious that when higher $\alpha$ is applied, lower outage probability can be achieved over the same feedback channel, which also holds when $\alpha>1$. 
The result shows that it is feasible to lower outage probability via asymmetric feedback detection compared with conventional symmetric feedback detection scheme.

\begin{figure}[!t]
\centering
\includegraphics[width=\linewidth]{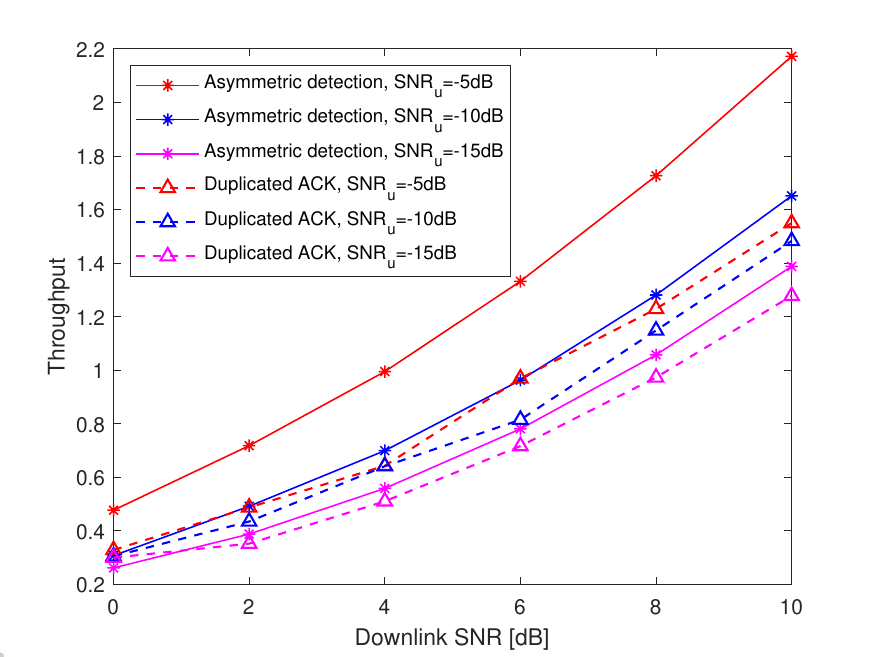}
\centering
\caption{The performance of HARQ process with asymmetric feedback detection compared with duplicated-ACK HARQ under the same outage probability constraints. 
}
\label{ASvs2ACK}
\end{figure}

In the absence of asymmetric detection, the transmitter might need to consider a block as delivered only if duplicated ACKs are observed to reduce the outage probability. Under the same outage probability limitation, i.e., $\epsilon=1\%$, we compare the case where the receiver starts sending subsequent blocks only after observing two ACKs with the case where asymmetric detection is applied. It is shown in Fig. \ref{ASvs2ACK} that asymmetric feedback detection outperforms the duplicated-ACK method under the constraints of outage probability. When the quality of the feedback channel is not very low, the throughput gain of asymmetric detection is more significant.

\begin{figure}[!t]
\centering
\includegraphics[width=\linewidth]{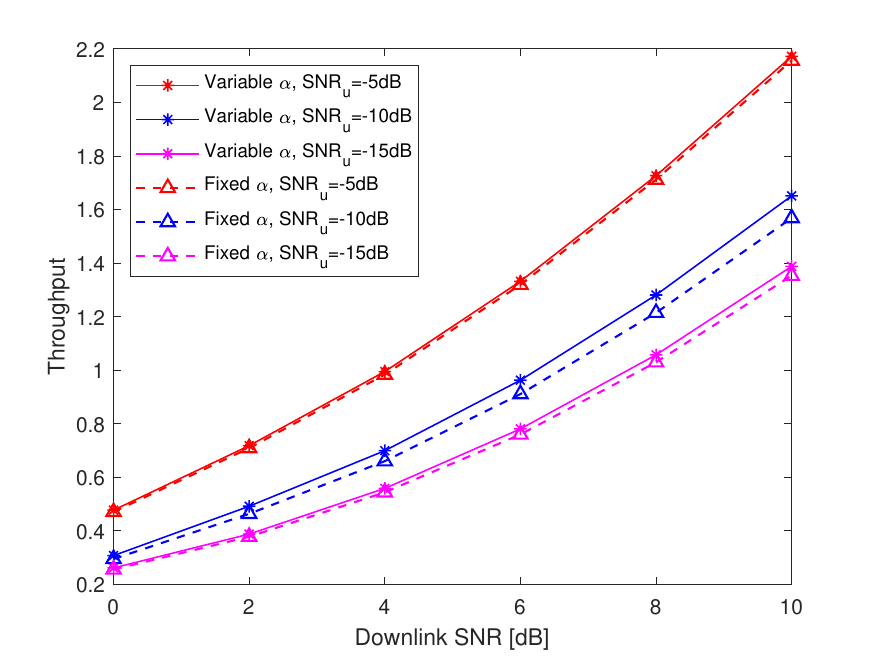}
\centering
\caption{The comparison of HARQ with fixed detection thresholds ($\alpha$) and variable thresholds when asymmetric feedback detection is applied.}
\label{Fixedalpha}
\end{figure}

We also compare the HARQ performance of fixed asymmetric thresholds and variable thresholds for feedback channel signal-to-noise ratio SNR$_u=-5,-10,-15$dB. The results in Fig.\ref{Fixedalpha} show that although not significant, a certain throughput gain can be obtained especially when SNR$_u$=-10dB.

\section{Conclusion}
In this paper, we analyse IR-HARQ performance where transmission attempts are carried over single user block-fading channel. We propose an asymmetric feedback detection scheme to protect NACK, and formulate a MDP model to optimize the throughout of the HARQ process when there is strict limitation on outage probability. Numerical results show that asymmetric feedback detection can significantly reduce the outage probability and achieve higher throughput than the other methods.

%\vspace{-.75em}
%\section*{Acknowledgment}
%\vspace{-.5em}
%This work was supported in part by the EPSRC through the Scalable Full Duplex Dense Wireless Networks (SENSE) grant EP/P003486/1, by the U.S. National Science Foundation
%under Grant  CNS-1836802, by the
%NSFC under grant 61871109, and by  the grants No. ZDSYS201707251409055, No. 2017ZT07X152, No. 2018B030338001, and No. 2018YFB1800800.

\vspace{-0.5em}
\bibliographystyle{IEEEtran}
\bibliography{IEEEabrv,MMM}

% that's all folks
\end{document}